\pdfoutput=1
\documentclass[11pt]{article}

\usepackage[utf8]{inputenc}
\usepackage[T1]{fontenc}
\usepackage{lmodern}
\usepackage{amsmath,amssymb,amsfonts,amsthm}
\usepackage{graphicx}
\usepackage{hyperref}
\usepackage{geometry}
\usepackage{mathrsfs}
\usepackage{enumitem}
\usepackage{bm}
\usepackage{listings}
\usepackage{float}
\numberwithin{equation}{section}

\newtheorem{theorem}{Theorem}[section]

\theoremstyle{definition}
\newtheorem{definition}[theorem]{Definition}
\theoremstyle{remark}

\title{Numerical Solution to the Riemann Problem\\
for a Liquid--Gas Two-phase Isentropic Flow Model}

\author{Abdul Rab\\
Department of Mathematics and Social Science\\
Sukkur IBA University, Pakistan\\
\texttt{abdulrab@live.com}} 

\date{}

\begin{document}

\maketitle

\begin{abstract}
A recently introduced two-phase flow model by Chun Shen is studied in this work. The model is derived to describe the dynamics of immersed water bubbles in liquid water as carrier. Several assumptions are made to obtain a reduced form of the mathematical model. The established model consists of nonlinear coupled PDEs in which the unknowns are the densities of the liquid and gas phases and the velocity of the liquid phase; these depend on space and time. For numerical purposes a one-dimensional space--time coordinate system $(x,t)$ is considered. Using the Python programming framework, several Riemann-type initial value problems for the two-phase flow model are solved numerically. A comparison of three finite-difference schemes is presented in order to examine their performance: the Lax--Friedrichs scheme, the Lax--Wendroff scheme, and the FORCE scheme. The FORCE scheme is total-variation diminishing (TVD) and monotone and does not create oscillations. As expected, the numerical solution of the Riemann problems consists of combinations of smooth profiles, shock waves, and rarefaction waves.
\end{abstract}

\section{Introduction}

Two-phase and multi-phase flows appear in a wide range of industrial applications, including petroleum extraction and transportation, genetic engineering, chemical engineering, and medical engineering. Because of the interaction between phases, these flows exhibit complex dynamical behaviour and a variety of flow regimes. Mathematically, two-phase flow models are typically described either by mixture theory or by averaging techniques, leading to nonlinear systems of conservation laws.

In this paper we consider a one-dimensional inviscid liquid--gas two-phase isentropic flow model of drift-flux type introduced by Shen . The model describes the motion of water with immersed gas bubbles under isentropic assumptions. After suitable simplifications, the governing equations can be written as a strictly hyperbolic $3\times3$ system of conservation laws. Our interest is in the numerical approximation of Riemann-type initial data and in the associated wave structure.

We review basic concepts from the theory of hyperbolic conservation laws, including weak solutions, entropy conditions, and the classification of waves into shocks, rarefactions and contact discontinuities. We then recall several classical finite-difference schemes (Lax--Friedrichs, Lax--Wendroff, and the FORCE scheme) and apply them to the liquid--gas model. The numerical results are obtained using Python and are used to illustrate the wave patterns predicted by the analytical theory.

The structure of the paper is as follows. In Section~\ref{sec:prelim} we collect preliminaries on scalar and system conservation laws as well as the numerical schemes that will be used. Section~\ref{sec:theory} briefly reviews the theory of nonlinear hyperbolic PDEs, including characteristic curves, rarefaction and shock waves, and Rankine--Hugoniot conditions. Section~\ref{sec:model} introduces the liquid--gas two-phase isentropic flow model, derives its characteristic structure, and discusses the associated Riemann problem. Section~\ref{sec:numerics} presents numerical simulations with the three schemes and a qualitative wave analysis of the results. We close with some conclusions and directions for future work.

\section{Preliminaries}\label{sec:prelim}

In this section we recall basic notions concerning scalar conservation laws and numerical schemes that will be used later.

\subsection{Definitions}

\begin{definition}[Scalar conservation law]
Let $G(x,t)$ be an unknown scalar function and $F:\mathbb{R}\to\mathbb{R}$ a flux function. The scalar conservation law is
\begin{equation}\label{eq:conslaw}
  \partial_t G(x,t) + \partial_x F(G(x,t)) = 0,\qquad x\in\mathbb{R},\ t>0.
\end{equation}
This PDE expresses conservation of a physical quantity $G$ in a control volume, with flow across the boundaries given by $F(G)$.
\end{definition}

\begin{definition}[Riemann problem]
A Riemann problem for the conservation law \eqref{eq:conslaw} consists of \eqref{eq:conslaw} together with piecewise constant initial data with a single discontinuity:
\begin{equation}
  G(x,0)=G_0(x)=
  \begin{cases}
    G_-, & x< x_0,\\[1ex]
    G_+, & x> x_0,
  \end{cases}
\end{equation}
where $G_-,G_+$ are constants.
\end{definition}

\begin{definition}[Characteristic curves]
For a sufficiently smooth solution $G(x,t)$ of \eqref{eq:conslaw}, a characteristic curve is a parametrized curve $t\mapsto x(t)$ along which $G$ is constant. Using the chain rule, the PDE can be written in quasi-linear form
\begin{equation}
  \partial_t G + a(G)\,\partial_x G = 0, \qquad a(G)=F'(G).
\end{equation}
Characteristics are then defined by the ODE
\begin{equation}
  \frac{dx}{dt}=a(G(x(t),t)),\qquad \frac{dG}{dt}=0.
\end{equation}
\end{definition}

\begin{definition}[Conservative numerical method]
Let $U_j^n$ denote a grid function approximating $G$ at $x_j$ and time $t^n$. A semi-discrete conservative finite-difference scheme for \eqref{eq:conslaw} has the form
\begin{equation}
  U_j^{n+1}= U_j^n - \frac{k}{h}\Big( \widehat F_{j+\frac12}^n - \widehat F_{j-\frac12}^n \Big),
\end{equation}
where $h$ is the spatial step, $k$ is the time step, and $\widehat F_{j\pm\frac12}^n$ are consistent numerical fluxes depending on neighbouring values $U_{j-p}^n,\dots,U_{j+q}^n$.
\end{definition}

\begin{definition}[Consistency]
A conservative scheme is consistent with the conservation law \eqref{eq:conslaw} if, for constant data $U_j^n \equiv \bar G$, the numerical flux reduces to the physical flux:
\begin{equation}
  \widehat F(\bar G,\dots,\bar G)=F(\bar G).
\end{equation}
\end{definition}

\begin{definition}[Weak solution]
A locally integrable function $G(x,t)$ is a weak solution of \eqref{eq:conslaw} with initial data $G_0$ if
\begin{equation}
  \int_0^\infty \int_{-\infty}^\infty \big( G\, \partial_t \phi + F(G)\,\partial_x \phi \big)\,dx\,dt
  = -\int_{-\infty}^\infty G_0(x)\,\phi(x,0)\,dx
\end{equation}
for all test functions $\phi\in C_c^\infty(\mathbb{R}\times[0,\infty))$.
\end{definition}

\begin{definition}[Convergent scheme]
A conservative and consistent numerical method is called convergent if, as $k,h\to0$ with the CFL ratio $k/h$ satisfying a suitable stability condition, the numerical solution converges (in an appropriate sense) to a weak solution of the conservation law.
\end{definition}

\begin{definition}[Linear scheme]
A linear scheme applied to a linear conservation law has the form
\begin{equation}
  U_j^{n+1} = \sum_{\ell=-p}^{q} c_\ell(\nu)\,U_{j+\ell}^n,\qquad \nu=\frac{k}{h},
\end{equation}
for coefficients $c_\ell(\nu)$ depending only on the CFL number.
\end{definition}

\begin{definition}[Monotone scheme]
A scheme of the form
\begin{equation}
  U_j^{n+1} = \mathcal{H}\big(U_{j-p}^n,\dots,U_{j+q}^n\big)
\end{equation}
is monotone if $\mathcal{H}$ is nondecreasing in each of its arguments.
\end{definition}

\begin{definition}[TVD scheme]
A scheme is total-variation diminishing (TVD) if the discrete total variation
\begin{equation}
  TV(U^n) = \sum_j \big| U_{j+1}^n - U_j^n \big|
\end{equation}
satisfies
\begin{equation}
  TV(U^{n+1}) \leq TV(U^n)\quad\text{for all }n.
\end{equation}
\end{definition}

\subsection{Some theorems}

We recall a few classical results.

\begin{theorem}[Oleinik entropy condition]
Suppose a piecewise smooth solution of \eqref{eq:conslaw} contains a single discontinuity with left and right states $G_-$ and $G_+$. Then $G(x,t)$ is an entropy solution if and only if
\begin{equation}
  \frac{F(G)-F(G_-)}{G-G_-} \;\geq\;
  \frac{F(G_+)-F(G_-)}{G_+-G_-} \;\geq\;
  \frac{F(G_+)-F(G)}{G_+-G}
\end{equation}
for all $G$ between $G_-$ and $G_+$.
\end{theorem}

If $F$ is convex, this condition simplifies.

\begin{theorem}
Assume $F$ is convex. Then a piecewise smooth solution with a single discontinuity is an entropy solution if and only if
\begin{equation}
  F'(G_-) \geq \frac{F(G_+)-F(G_-)}{G_+-G_-} \geq F'(G_+).
\end{equation}
\end{theorem}

\begin{theorem}[Godunov]
A linear conservative scheme is monotone if and only if it is TVD. Moreover, any linear TVD scheme is at most first-order accurate.
\end{theorem}

\subsection{Numerical schemes}

We briefly recall the Lax--Wendroff, Lax--Friedrichs and FORCE schemes.

\subsubsection{Lax--Wendroff scheme (Richtmyer form)}

For a conservation law $G_t + F(G)_x=0$, the two-step Richtmyer version of the Lax--Wendroff scheme reads
\begin{align}
  G_{i+\frac12}^{n+\frac12} &= \frac12\big(G_{i+1}^n+G_i^n\big)
  - \frac{\Delta t}{2\Delta x}\big( F(G_{i+1}^n)-F(G_i^n) \big),\\
  G_i^{n+1} &= G_i^n - \frac{\Delta t}{\Delta x}
  \Big( F\big(G_{i+\frac12}^{n+\frac12}\big) - F\big(G_{i-\frac12}^{n+\frac12}\big) \Big).
\end{align}

\subsubsection{Lax--Friedrichs scheme}

The Lax--Friedrichs method is given by
\begin{equation}
  G_i^{n+1} = \frac12\big( G_{i+1}^n + G_{i-1}^n \big)
  - \frac{\Delta t}{2\Delta x}\Big( F(G_{i+1}^n) - F(G_{i-1}^n) \Big).
\end{equation}
It is first-order accurate and strongly diffusive, but robust and easy to implement.

\subsubsection{FORCE scheme}

The first-order centered (FORCE) scheme combines the Lax--Friedrichs and Lax--Wendroff fluxes. The numerical flux at $x_{i-\frac12}$ is
\begin{equation}
  \widehat F_{i-\frac12}^{\mathrm{FORCE}}
  = \frac12\big( \widehat F_{i-\frac12}^{\mathrm{LF}} + \widehat F_{i-\frac12}^{\mathrm{LW}} \big),
\end{equation}
where
\begin{align}
  \widehat F_{i-\frac12}^{\mathrm{LF}}
    &= \frac12\big( F(G_{i-1}) + F(G_i) \big)
       - \frac12\frac{\Delta x}{\Delta t}(G_i-G_{i-1}),\\
  G_{i-\frac12}^{\mathrm{LW}}
    &= \frac12\big( G_{i-1} + G_i\big)
       - \frac12\frac{\Delta t}{\Delta x}\big( F(G_i)-F(G_{i-1})\big),\\
  \widehat F_{i-\frac12}^{\mathrm{LW}}
    &= F\big(G_{i-\frac12}^{\mathrm{LW}}\big),
\end{align}
and similar expressions hold at $x_{i+\frac12}$.

\section{Theory of Hyperbolic Nonlinear PDEs}\label{sec:theory}

\subsection{Nonlinear scalar conservation laws}

We consider again the nonlinear conservation law
\begin{equation}
  \partial_t G + \partial_x F(G)=0,
\end{equation}
with $F\in C^2(\mathbb{R})$. The characteristic speed is $a(G)=F'(G)$ and we distinguish:
\begin{itemize}[leftmargin=1.8em]
  \item convex flux: $F''(G)\ge0$ for all $G$,
  \item concave flux: $F''(G)\le0$ for all $G$,
  \item general flux: $F''$ changes sign.
\end{itemize}

\paragraph{Example (Burgers' equation).}
The inviscid Burgers equation
\begin{equation}
  \partial_t G + \partial_x\Big( \frac12 G^2 \Big)=0
\end{equation}
has flux $F(G)=\tfrac12 G^2$. Then $F'(G)=G$ and $F''(G)=1\ge0$, so the flux is convex.

\subsection{Self-similar rarefaction waves}

For Riemann data
\begin{equation}
  G(x,0)=
  \begin{cases}
    G_-, & x<x_0,\\
    G_+, & x>x_0,
  \end{cases}
\end{equation}
with convex flux and $a(G_-)\le a(G_+)$, the solution may contain a centered rarefaction wave. Assuming self-similarity $G(x,t)=G(\xi)$ with $\xi=(x-x_0)/t$ and substituting into the PDE yields
\begin{equation}
  ( -\xi + a(G(\xi)) )\,G'(\xi)=0,
\end{equation}
so in the continuous part of the rarefaction
\begin{equation}
  a(G(\xi)) = \xi.
\end{equation}
The complete solution is
\begin{equation}
  G(x,t)=
  \begin{cases}
    G_-, & \dfrac{x-x_0}{t}<a(G_-),\\[1ex]
    G(\xi),\ \ a(G_-)\le\xi\le a(G_+),\\[1ex]
    G_+, & \dfrac{x-x_0}{t}>a(G_+),
  \end{cases}
\end{equation}
where $G(\xi)$ is obtained by solving $a(G)=\xi$.

\subsection{Shock waves and Rankine--Hugoniot condition}

When characteristics intersect, the classical solution breaks down and shocks form. Consider a discontinuity moving with speed $s$ and left/right states $G_-$, $G_+$. Integrating the conservation law over a space--time control volume and letting its width shrink to the shock yields the Rankine--Hugoniot condition
\begin{equation}
  s(G_+-G_-) = F(G_+)-F(G_-).
\end{equation}

\paragraph{Example (Burgers' equation).}
For $F(G)=\tfrac12 G^2$, the shock speed is
\begin{equation}
  s = \frac{\tfrac12 G_+^2 - \tfrac12 G_-^2}{G_+-G_-}
    = \frac12(G_+ + G_-).
\end{equation}

\subsection{Systems of conservation laws}

A general $m\times m$ system of conservation laws in one space dimension has the form
\begin{equation}\label{eq:system}
  \partial_t \mathbf{G}(t,x) + \partial_x \mathbf{F}(\mathbf{G}(t,x)) = 0,
\end{equation}
where $\mathbf{G}=(G_1,\dots,G_m)^\top$ is the vector of conserved variables and $\mathbf{F}=(F_1,\dots,F_m)^\top$ the flux vector. Under suitable smoothness, \eqref{eq:system} can be written in quasi-linear form
\begin{equation}
  \partial_t \mathbf{G} + A(\mathbf{G})\,\partial_x \mathbf{G} = 0,\qquad
  A(\mathbf{G}) = D\mathbf{F}(\mathbf{G}).
\end{equation}
The eigenvalues $\lambda_i(\mathbf{G})$ and right eigenvectors $\mathbf{r}_i(\mathbf{G})$ of $A(\mathbf{G})$ determine the characteristic speeds and fields. The system is \emph{hyperbolic} if all eigenvalues are real and $A(\mathbf{G})$ has a complete set of eigenvectors; it is \emph{strictly hyperbolic} if the eigenvalues are also distinct.

A characteristic field $i$ is called \emph{genuinely nonlinear} if
\begin{equation}
 \nabla_{\mathbf{G}} \lambda_i(\mathbf{G})\cdot \mathbf{r}_i(\mathbf{G}) \neq 0
 \quad\text{for all }\mathbf{G},
\end{equation}
and \emph{linearly degenerate} if this quantity is identically zero. Genuinely nonlinear fields generate shock and rarefaction waves; linearly degenerate fields generate contact discontinuities.

\section{Liquid--Gas Two-phase Isentropic Flow Model}\label{sec:model}

\subsection{Model equations}

Following Shen Model, we consider the reduced drift-flux type model
\begin{equation}\label{eq:model}
\begin{cases}
 m_t + (mu)_x = 0,\\[0.4ex]
 n_t + (nu)_x = 0,\\[0.4ex]
 (nu)_t + (nu^2 + k(m+n)^\gamma)_x = 0,
\end{cases}
\end{equation}
where $m=m(x,t)$ and $n=n(x,t)$ denote the masses of the gas and liquid phases, respectively, $u=u(x,t)$ is the common velocity, and $k>0$, $\gamma>1$ are physical parameters. The model is derived from the more general two-phase flow system
\begin{equation}
\begin{cases}
 (\alpha_g\rho_g)_t + (\alpha_g\rho_g u_g)_x = 0,\\
 (\alpha_\ell\rho_\ell)_t + (\alpha_\ell\rho_\ell u_\ell)_x = 0,\\
 (\alpha_g\rho_g u_g + \alpha_\ell\rho_\ell u_\ell)_t
 + (\alpha_g\rho_g u_g^2 + \alpha_\ell\rho_\ell u_\ell^2)_x = 0,
\end{cases}
\end{equation}
under the identifications $m=\alpha_g\rho_g$, $n=\alpha_\ell\rho_\ell$, and suitable closure assumptions. The volume fractions $\alpha_g,\alpha_\ell\in[0,1]$ satisfy $\alpha_g+\alpha_\ell=1$.

We consider the Riemann problem
\begin{equation}\label{eq:Riemann}
  (m,n,u)(x,0) =
  \begin{cases}
   (m_-,n_-,u_-), & x<0,\\
   (m_+,n_+,u_+), & x>0,
  \end{cases}
\end{equation}
and seek self-similar weak solutions depending on $\xi=x/t$.

\subsection{Characteristic structure}

Writing \eqref{eq:model} in quasi-linear form yields
\begin{equation}
  A(m,n,u)
  \begin{pmatrix}
   m\\ n\\ u
  \end{pmatrix}_t
  +
  B(m,n,u)
  \begin{pmatrix}
   m\\ n\\ u
  \end{pmatrix}_x
  = \mathbf{0},
\end{equation}
where, after rearrangement, we can identify the Jacobian matrix associated with the flux in \eqref{eq:model}. A convenient way is to consider the flux
\begin{equation}
  \mathbf{F}(m,n,u) =
  \begin{pmatrix}
   mu\\ nu\\ nu^2 + k(m+n)^\gamma
  \end{pmatrix}
\end{equation}
and compute its Jacobian $A(\mathbf{G})=D\mathbf{F}(\mathbf{G})$. This leads to the characteristic polynomial
\begin{equation}
  (\lambda-u)\big( n(\lambda-u)^2 - k\gamma(m+n)^{\gamma-1}\big)=0,
\end{equation}
with eigenvalues
\begin{equation}
 \lambda_1 = u - \sqrt{\frac{k\gamma(m+n)^\gamma}{n}},\qquad
 \lambda_2 = u,\qquad
 \lambda_3 = u + \sqrt{\frac{k\gamma(m+n)^\gamma}{n}}.
\end{equation}
The corresponding right eigenvectors can be chosen as
\begin{align}
 \mathbf{r}_1 &= (m,n,-\sqrt{\tfrac{k\gamma(m+n)^\gamma}{n}})^\top,\\
 \mathbf{r}_2 &= (1,-1,0)^\top,\\
 \mathbf{r}_3 &= (m,n,\sqrt{\tfrac{k\gamma(m+n)^\gamma}{n}})^\top.
\end{align}
For $k>0$ the system is strictly hyperbolic. Computing
\begin{align}
 \nabla \lambda_1 \cdot \mathbf{r}_1 &= -\frac{\gamma+1}{2}
   \sqrt{\frac{k\gamma(m+n)^\gamma}{n}},\\
 \nabla \lambda_2 \cdot \mathbf{r}_2 &= 0,\\
 \nabla \lambda_3 \cdot \mathbf{r}_3 &= \frac{\gamma+1}{2}
   \sqrt{\frac{k\gamma(m+n)^\gamma}{n}},
\end{align}
shows that fields 1 and 3 are genuinely nonlinear, while field~2 is linearly degenerate. Thus, waves associated with $\lambda_1$ and $\lambda_3$ are shock or rarefaction waves, whereas waves associated with $\lambda_2$ are contact discontinuities.

\section{Numerical Simulation}\label{sec:numerics}

We now apply the schemes from Section~\ref{sec:prelim} to the model \eqref{eq:model} with Riemann initial data.

\subsection{Lax--Friedrichs scheme}

We discretize \eqref{eq:model} on a uniform grid $x_i=i\Delta x$, $t^n=n\Delta t$ using the Lax--Friedrichs method componentwise. Two sets of Riemann data are considered.

\paragraph{Initial data set 1.}
\begin{equation}
 (m,n,u)(x,0) =
 \begin{cases}
   (1.5,1.8,2.0), & x<0,\\
   (1.2,1.0,2.5), & x>0.
 \end{cases}
\end{equation}

\paragraph{Initial data set 2.}
\begin{equation}
 (m,n,u)(x,0) =
 \begin{cases}
   (0.8,1.0,1.2), & x<0,\\
   (1.0,0.8,1.0), & x>0.
 \end{cases}
\end{equation}

The numerical solutions are computed using Python 3. For initial data set~1 we display the profiles of $m$, $n$ and $u$ at $t=0.17$; for initial data set~2 we display the same quantities at $t=0.25$.

\begin{figure}[H]
 \centering
 \includegraphics[width=0.8\textwidth]{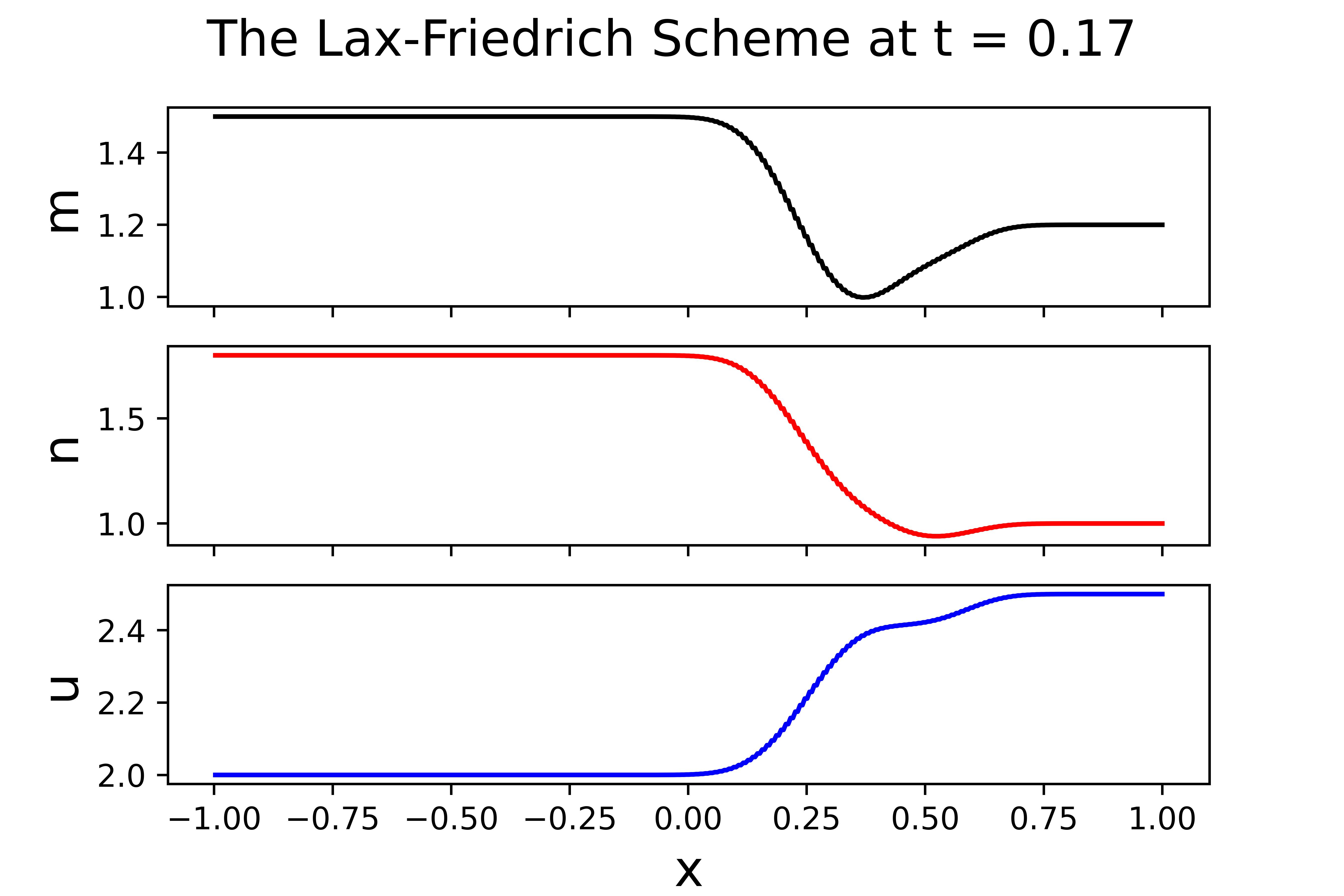}
 \caption{Lax--Friedrichs scheme at $t=0.17$ for initial data set~1.}
 \label{fig:LF1}
\end{figure}

\begin{figure}[H]
 \centering
 \includegraphics[width=0.8\textwidth]{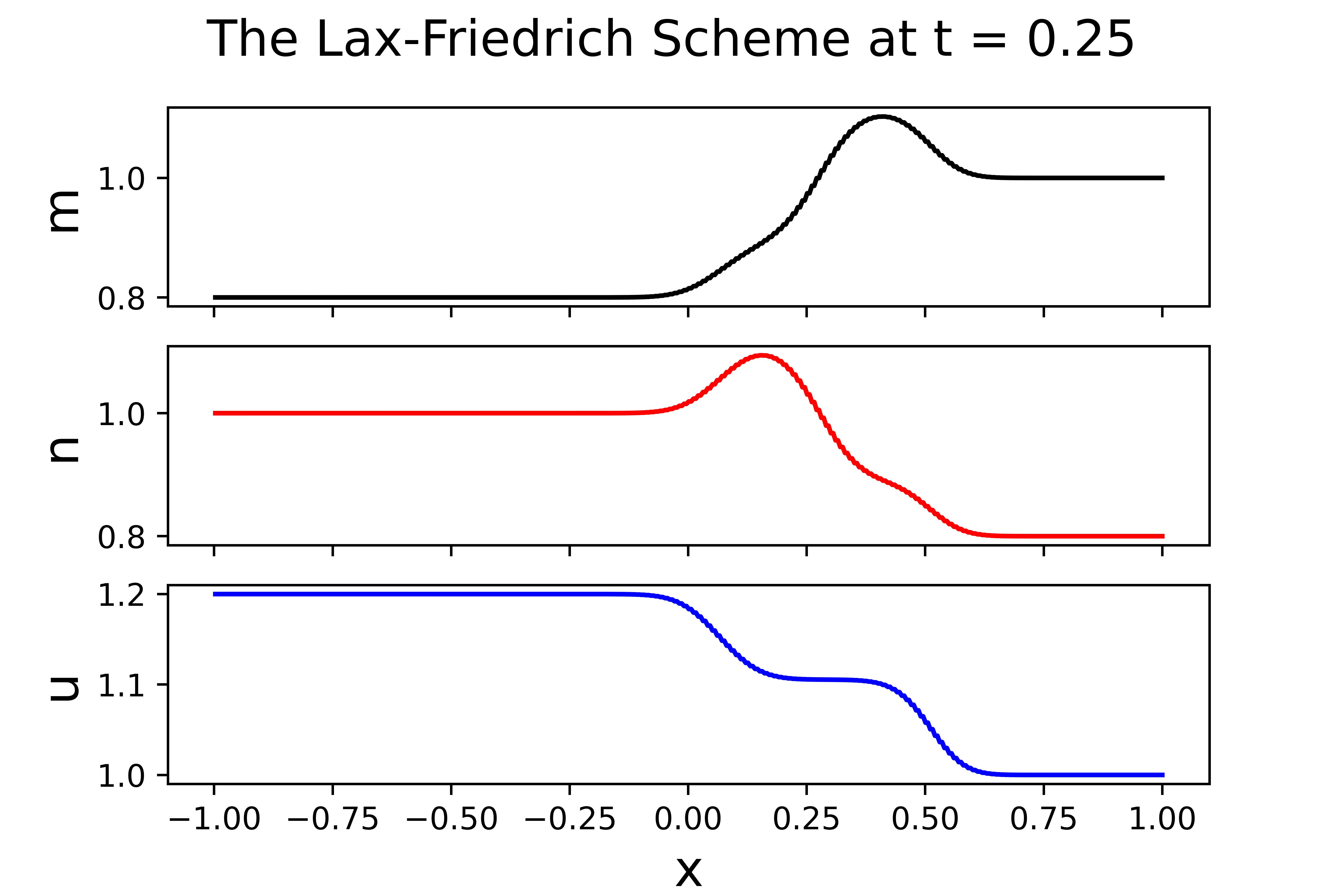}
 \caption{Lax--Friedrichs scheme at $t=0.25$ for initial data set~2.}
 \label{fig:LF2}
\end{figure}

(Here and below, the actual figure files should be supplied as external graphics when submitting to arXiv.)

\subsection{Lax--Wendroff scheme}

The Lax--Wendroff (Richtmyer) scheme is also applied to \eqref{eq:model} with the same Riemann data. Again, we show results at $t=0.15$ for data set~1 and $t=0.25$ for data set~2.

\begin{figure}[H]
 \centering
 \includegraphics[width=0.8\textwidth]{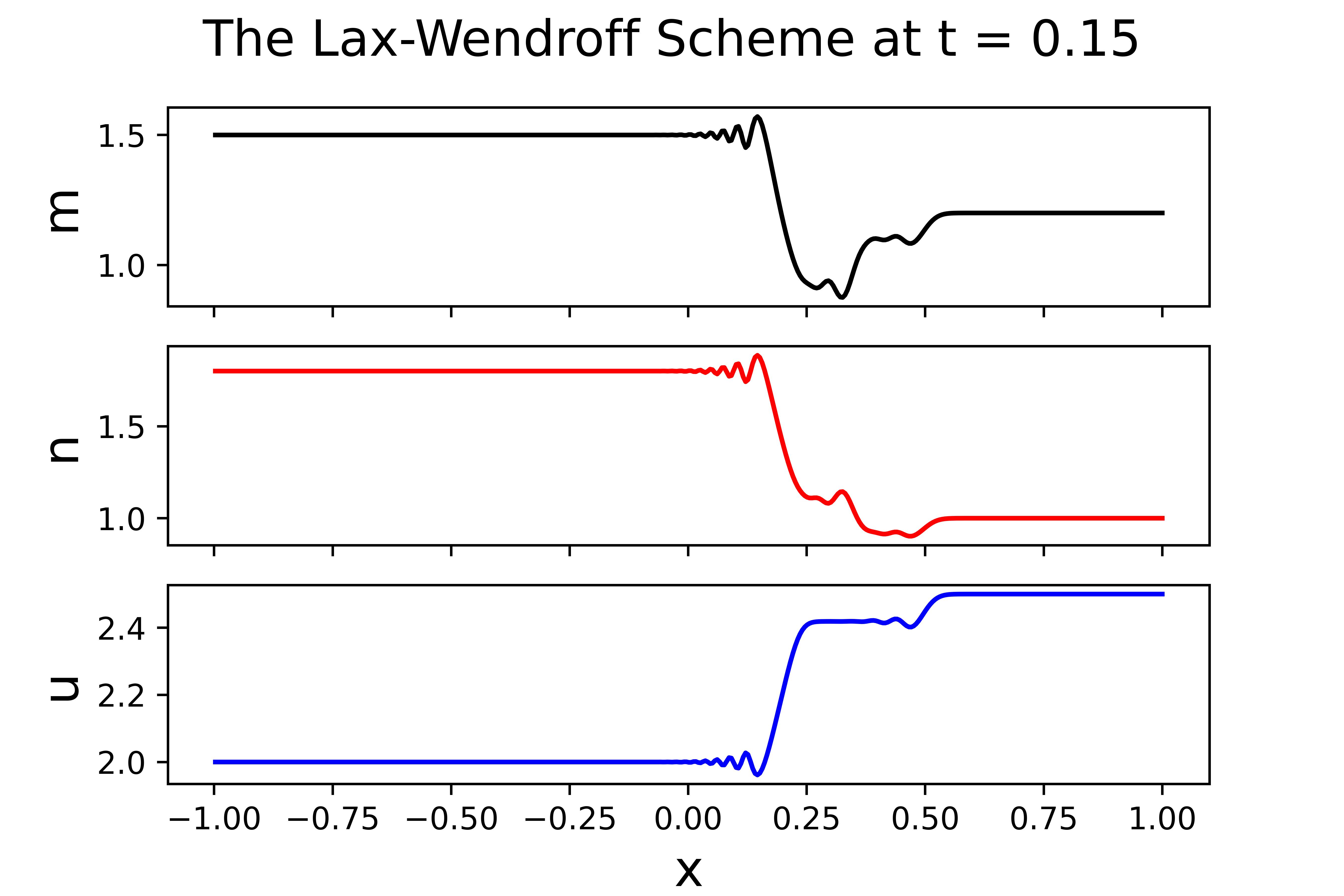}
 \caption{Lax--Wendroff scheme at $t=0.15$ for initial data set~1.}
 \label{fig:LW1}
\end{figure}

\begin{figure}[H]
 \centering
 \includegraphics[width=0.8\textwidth]{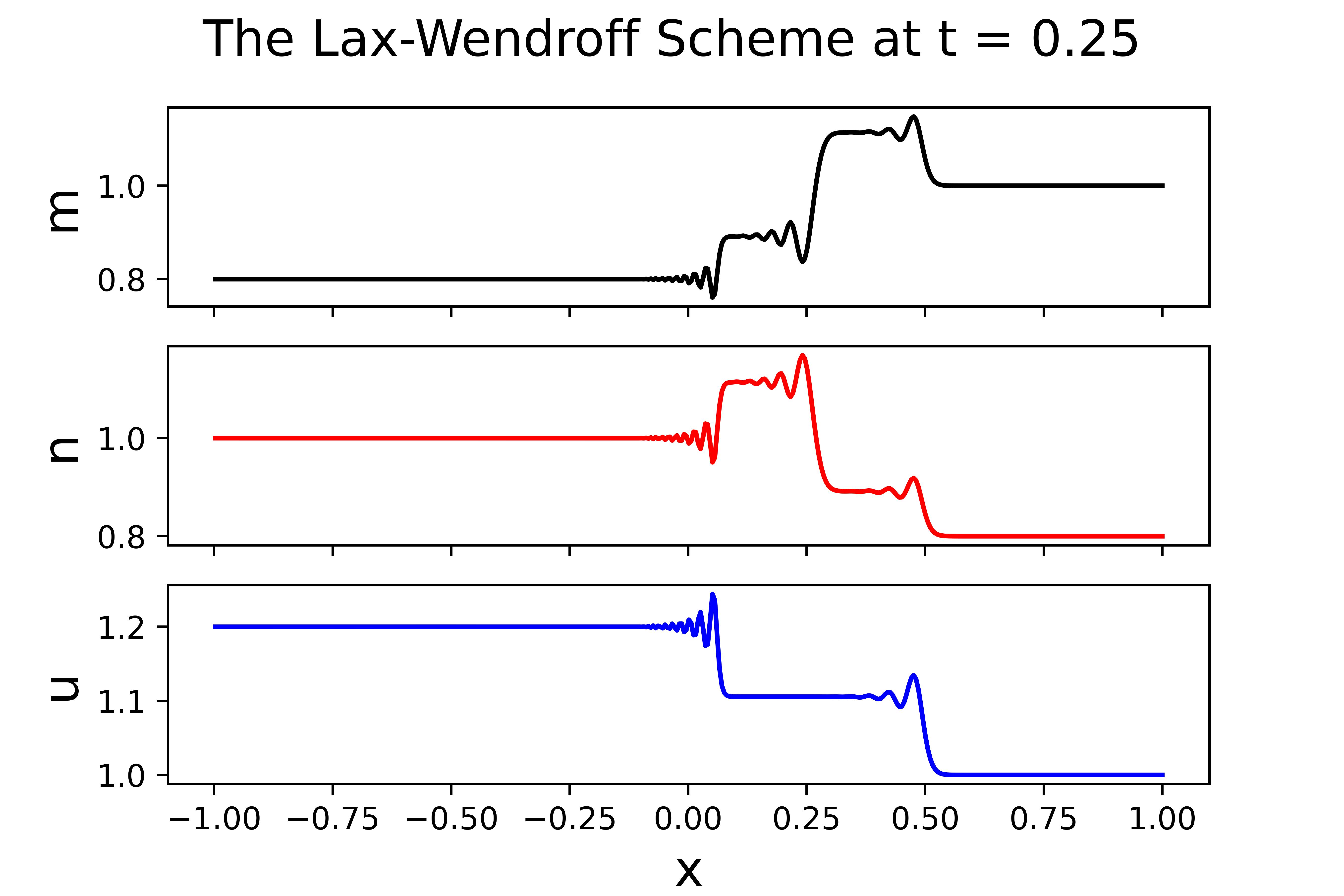}
 \caption{Lax--Wendroff scheme at $t=0.25$ for initial data set~2.}
 \label{fig:LW2}
\end{figure}

\subsection{FORCE scheme}

Finally, we apply the FORCE scheme. As before, we consider both sets of Riemann data.

\begin{figure}[H]
 \centering
 \includegraphics[width=0.8\textwidth]{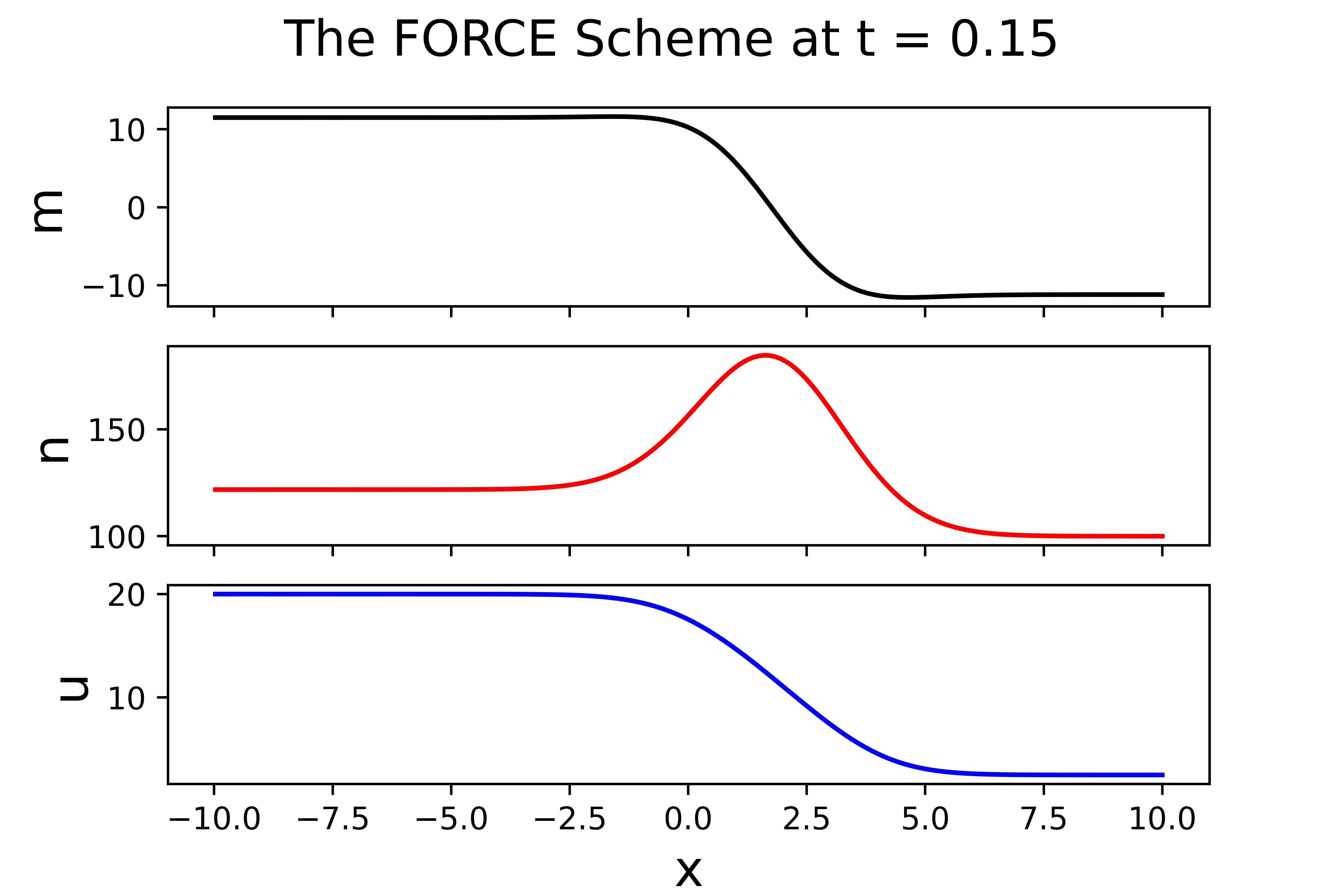}
 \caption{FORCE scheme at $t=0.15$ for initial data set~1.}
 \label{fig:FORCE1}
\end{figure}

\begin{figure}[H]
 \centering
 \includegraphics[width=0.8\textwidth]{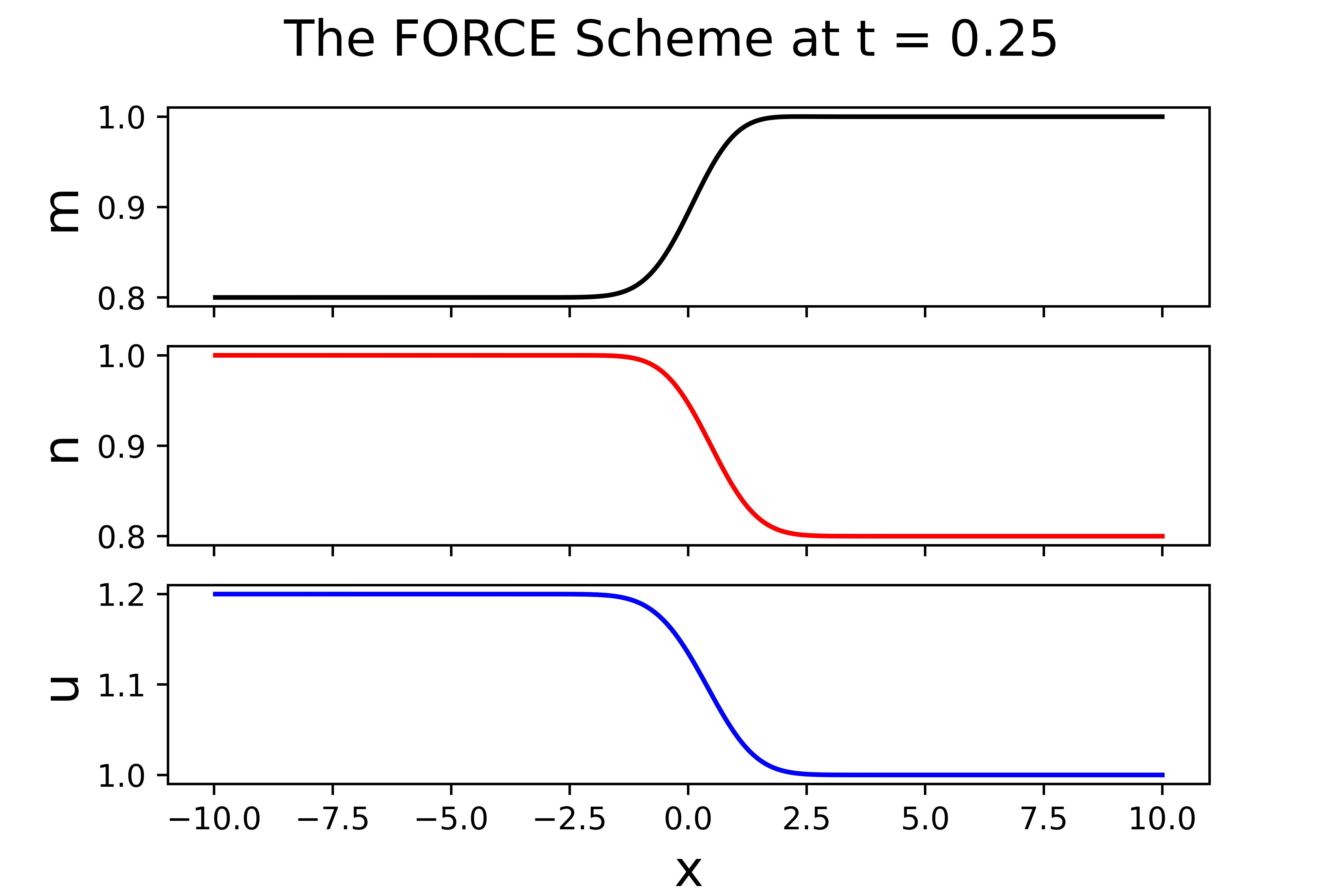}
 \caption{FORCE scheme at $t=0.25$ for initial data set~2.}
 \label{fig:FORCE2}
\end{figure}

\subsection{Wave analysis of numerical solutions}

We briefly describe the observed wave patterns.

For initial data set~1 with the Lax--Friedrichs scheme (Figure~\ref{fig:LF1}), the variables $m$ and $n$ exhibit two rarefaction waves: one moving to the left and one to the right. For the velocity $u$, two rarefaction waves moving to the right are visible.

For initial data set~2 (Figure~\ref{fig:LF2}), $m$ and $n$ again show two rarefaction waves, but now the left rarefaction moves to the right and the right rarefaction to the left. For $u$, two rarefaction waves propagate to the left.

The Lax--Wendroff results (Figures~\ref{fig:LW1}--\ref{fig:LW2}) resolve more structure. For instance, for data set~1 the solutions for $m$ and $n$ consist of three elementary waves: a right-moving shock, a contact discontinuity, and a right-moving rarefaction. The velocity $u$ exhibits one shock and one rarefaction wave, both moving to the right. For data set~2, $m$ combines a left-moving shock, a contact discontinuity, and a left-moving rarefaction, while $n$ shows a similar pattern with different ordering. The velocity $u$ contains two shocks.

The FORCE scheme produces smooth profiles even though the initial data are discontinuous. However, it fails to capture all elementary waves as clearly as the Lax--Wendroff scheme. In particular, the TVD and monotone nature of FORCE suppresses oscillations but also smears out some wave structures. In our experiments the Lax--Wendroff scheme gives a more accurate representation of the underlying Riemann solutions for this problem.

\section{Conclusions and Future Work}

We have studied a reduced drift-flux model for liquid--gas two-phase isentropic flow and solved Riemann problems numerically using three finite-difference schemes. The analytical structure of the model was examined via its characteristic fields, revealing two genuinely nonlinear fields and one linearly degenerate field. Numerical simulations using Lax--Friedrichs, Lax--Wendroff, and FORCE schemes show combinations of rarefaction waves, shock waves, and contact discontinuities in agreement with the theoretical predictions and with earlier work by Shen.

Among the tested schemes, the Lax--Wendroff method provided the clearest resolution of elementary waves, while the FORCE scheme produced smoother but more diffusive solutions and did not capture all features. Future work includes the implementation of higher-order schemes, such as Roe-type approximate Riemann solvers, MUSCL reconstructions, and high-order finite-volume schemes with ENO/WENO reconstructions, in order to obtain more accurate and less diffusive approximations of the two-phase flow model.

\section*{Acknowledgements}

(The acknowledgements from the thesis can be inserted here if desired.)

\appendix

\section{Python Code}\label{app:code}

In this appendix we include (or reference) the Python code used to generate the numerical results. For arXiv submission it is usually more convenient to provide the code as separate source files, but short scripts can be embedded using the \texttt{listings} package.

\begin{lstlisting}[language=Python, basicstyle=\ttfamily\small]
"""
Created on Sat Oct 1 02:23:17 2022

@author: Abdul Rab Chachar
"""

import os, sys
import numpy as np
import matplotlib.pyplot as plt
from numpy import *
from matplotlib import rc


# Parameters
k      = 0.6
CFL    = 0.50                    
gamma  = 0.8                     
ncells = 400                     # Number of cells
xx_initial = -1.0; xx_final = 1.0         # Limits of computational domain 
step_size = (xx_final-xx_initial)/ncells        # Step size of spatial variable
nx = ncells+1                    # Number of points
domain = np.linspace(xx_initial+step_size/2.,xx_final,nx) # Computational Mesh of spatial variable.

# Pre allocation of Initial Conditions
m0 = zeros(nx)
n0 = zeros(nx)
u0 = zeros(nx)
halfcells = int(ncells/2)

IC = 1 # 6 Initial Conditoins cases are following
if IC == 1:
    print ("Initial Data Set 1")
    m0[:halfcells] = 1.5  ; m0[halfcells:] = 1.2;
    n0[:halfcells] = 1.8  ; n0[halfcells:] = 1.0;
    u0[:halfcells] = 2.0  ; u0[halfcells:] = 2.5;
    tEnd = 0.15;
elif IC == 2:
    print ("Initial Data Set 2")
    m0[:halfcells] = 0.8  ; m0[halfcells:] = 1.0;
    n0[:halfcells] = 1.0  ; n0[halfcells:] = 0.8;
    u0[:halfcells] = 1.2  ; u0[halfcells:] = 1.0;
    tEnd = 0.25;
elif IC == 3:
    print ("Initial Data Set 3")
    m0[:halfcells] = 2.0  ; m0[halfcells:] = 3.0;
    n0[:halfcells] = 2.0  ; n0[halfcells:] = 1.0;
    u0[:halfcells] = 2.5  ; u0[halfcells:] = 0.5;
    tEnd = 0.25;
elif IC == 4:
    print ("Initial Data Set 4")
    m0[:halfcells] = 3.2  ; m0[halfcells:] = 3.3;
    n0[:halfcells] = 2.5  ; n0[halfcells:] = 2.2;
    u0[:halfcells] = 2.9  ; u0[halfcells:] = 0.8;
    tEnd = 0.25;


q  = np.array([m0,n0,n0*u0])   # Vector of conserved variables

# Solver loop
t  = 0
it = 0
lamda3 = max(u0+((k*gamma*(m0+n0)**gamma)/n0)**0.5)
dt=CFL*step_size/lamda3   # Using the system's largest eigenvalue



# Plot the Initial Data Set 
fig, axs = plt.subplots(3,sharex=True)
fig.suptitle('The Lax-Wendroff Scheme - Initial Data Set '+str(IC)+'',fontsize=16)
axs[0].plot(domain, q[0],'k-')
axs[0].set_ylabel('m',fontsize=16)
axs[1].plot(domain, q[1],'r-')
axs[1].set_ylabel('n',fontsize=16)
axs[2].plot(domain, q[2],'b-')
axs[2].set_ylabel('u',fontsize=16)
axs[2].set_xlabel('x',fontsize=16)
fig.savefig("Lax_Wendroff_Scheme_Initial_"+str(IC)+".jpeg", dpi=900)
plt.show()


while t < tEnd:
    
    # Creating the copy of the conserved variable.
    Q = q.copy();
    
    # Primary variables
    m=Q[0];
    n=Q[1];
    u=Q[2]/n;
    
    # Flux vector of conserved variable
    F0 = np.array(m*u)
    F1 = np.array(n*u)
    F2 = np.array(n*u**2 +k*(m+n)**gamma)
    flux=np.array([ F0, F1, F2 ]);
    
    qm  = np.roll(Q, 1);
    qp  = np.roll(Q,-1);
    fm  = np.roll(flux, 1);
    fp  = np.roll(flux,-1);

    qpHalf = (qp+Q)/2. - dt/(2.*step_size)*(fp-flux)
    qmHalf = (qm+Q)/2. - dt/(2.*step_size)*(flux-fm)
    
    m=qpHalf[0];
    n=qpHalf[1];
    u=qpHalf[2]/n;
    
    F0 = np.array(m*u);
    F1 = np.array(n*u);
    F2 = np.array(n*u**2+k*(m+n)**gamma);
    FqpHalf=np.array([ F0, F1, F2 ]);
    
    m=qmHalf[0];
    n=qmHalf[1];
    u=qmHalf[2]/n;
    
    F0 = np.array(m*u);
    F1 = np.array(n*u);
    F2 = np.array(n*u**2+k*(m+n)**gamma);
    FqmHalf=np.array([ F0, F1, F2 ]);
    
    dF = FqpHalf - FqmHalf;
    
    q = Q-dt/step_size*dF;
    
    # Applying the Boundary Conditions
    q[:,0]=Q[:,0]; q[:,-1]=Q[:,-1]; 

    # Compute primary variables
    m=q[0];
    n=q[1];
    u=q[2]/n;

    # Update time step with respect eigenvalue
    lamda3 = max(u+((k*gamma*(m+n)**gamma)/n)**0.5);
    dt=CFL*step_size/lamda3;
    
    # Update time 
    t=t+dt;
    

fig, axs = plt.subplots(3,sharex=True)
fig.suptitle('The Lax-Wendroff Scheme at t = '+str(round(t,2))+'',fontsize=16)
axs[0].plot(domain,m,'k-')
axs[0].set_ylabel('m',fontsize=16)
axs[1].plot(domain, n,'r-')
axs[1].set_ylabel('n',fontsize=16)
axs[2].plot(domain, u,'b-')
axs[2].set_ylabel('u',fontsize=16)
axs[2].set_xlabel('x',fontsize=16)
fig.savefig("Lax_Wendroff_Scheme_Final_"+str(IC)+".jpeg", dpi=900)
plt.show()
--------------------------------
"""
Created on Sat Oct 5 22:34:08 2022

@author: Abdul Rab Chachar
"""

import os, sys
import numpy as np
import matplotlib.pyplot as plt
from numpy import *
from matplotlib import rc


# Parameters
k      = 0.6
CFL    = 0.50                    
gamma  = 0.8                     
ncells = 400                     # Number of cells
xx_initial = -1.0; xx_final = 1.0         # Limits of computational domain 
step_size = (xx_final-xx_initial)/ncells        # Step size of spatial variable
nx = ncells+1                    # Number of points
domain = np.linspace(xx_initial+step_size/2.,xx_final,nx) # Computational Mesh of spatial variable.

# Pre allocation of Initial Conditions
m0 = zeros(nx)
n0 = zeros(nx)
u0 = zeros(nx)
halfcells = int(ncells/2)

IC = 1 # 6 Initial Conditoins cases are following
if IC == 1:
    print ("Initial Data Set 1")
    m0[:halfcells] = 1.5  ; m0[halfcells:] = 1.2;
    n0[:halfcells] = 1.8  ; n0[halfcells:] = 1.0;
    u0[:halfcells] = 2.0  ; u0[halfcells:] = 2.5;
    tEnd = 0.15;
elif IC == 2:
    print ("Initial Data Set 2")
    m0[:halfcells] = 0.8  ; m0[halfcells:] = 1.0;
    n0[:halfcells] = 1.0  ; n0[halfcells:] = 0.8;
    u0[:halfcells] = 1.2  ; u0[halfcells:] = 1.0;
    tEnd = 0.25;
elif IC == 3:
    print ("Initial Data Set 3")
    m0[:halfcells] = 2.0  ; m0[halfcells:] = 3.0;
    n0[:halfcells] = 2.0  ; n0[halfcells:] = 1.0;
    u0[:halfcells] = 2.5  ; u0[halfcells:] = 0.5;
    tEnd = 0.25;
elif IC == 4:
    print ("Initial Data Set 4")
    m0[:halfcells] = 3.2  ; m0[halfcells:] = 3.3;
    n0[:halfcells] = 2.5  ; n0[halfcells:] = 2.2;
    u0[:halfcells] = 2.9  ; u0[halfcells:] = 0.8;
    tEnd = 0.25;
q  = np.array([m0,n0,n0*u0])   # Vector of conserved variables

# Solver loop starts here
t  = 0 # Initial Time
lamda3 = max(u0+((k*gamma*(m0+n0)**gamma)/n0)**0.5)  # Largest engenvalue of the system
dt=CFL*step_size/lamda3   # Using the system's largest eigenvalue

# Plot the Initial Data Set 
fig, axs = plt.subplots(3,sharex=True)
fig.suptitle('Initial Data Set '+str(IC)+'',fontsize=16)
axs[0].plot(domain, q[0],'k-')
axs[0].set_ylabel('m',fontsize=16)
axs[1].plot(domain, q[1],'r-')
axs[1].set_ylabel('n',fontsize=16)
axs[2].plot(domain, q[2],'b-')
axs[2].set_ylabel('u',fontsize=16)
axs[2].set_xlabel('x',fontsize=16)
fig.savefig("Initial_Data_Set_"+str(IC)+".jpeg", dpi=900)
plt.show()



while t < tEnd:
    Q = q.copy();

    # Primary variables
    m=Q[0];
    n=Q[1];
    u=Q[2]/n;

    # Flux vector of conserved properties
    F0 = np.array(m*u);
    F1 = np.array(n*u);
    F2 = np.array(n*u**2 +k*(m+n)**gamma);
    flux=np.array([ F0, F1, F2 ]);
    qm  = np.roll(Q, 1);
    qp  = np.roll(Q,-1);
    fm  = np.roll(flux, 1);
    fp  = np.roll(flux,-1);
    q = (qp+qm)/2. - dt/(2.*step_size)*(fp-fm)
    # Compute primary variables
    m=q[0];
    n=q[1];
    u=q[2]/n;
    q[:,0]=Q[:,0]; q[:,-1]=Q[:,-1]; # Boundary Conditions
    # Update/correct time step
    lamda3 = max(u+((k*gamma*(m+n)**gamma)/n)**0.5)
    dt=CFL*step_size/lamda3;
    # Update time and iteration counter
    t=t+dt;
fig, axs = plt.subplots(3,sharex=True)
fig.suptitle('The Lax-Friedrich Scheme at t = '+str(round(t,2))+'',fontsize=16)
axs[0].plot(domain,m,'k-')
axs[0].set_ylabel('m',fontsize=16)
axs[1].plot(domain, n,'r-')
axs[1].set_ylabel('n',fontsize=16)
axs[2].plot(domain, u,'b-')
axs[2].set_ylabel('u',fontsize=16)
axs[2].set_xlabel('x',fontsize=16)
fig.savefig("Lax_Friedrich_Scheme_Final_"+str(IC)+".jpeg", dpi=900)
plt.show()
------------------------------
"""
Created on Sat Oct 8 20:33:33 2022

@author: Abdul Rab Chachar
"""

import os, sys
import numpy as np
import matplotlib.pyplot as plt
from numpy import *
from matplotlib import rc


# Parameters
k      = 0.6
CFL    = 0.50                    
gamma  = 0.8                     
ncells = 400                     # Number of cells
xx_initial = -1.0; xx_final = 1.0         # Limits of computational domain 
step_size = (xx_final-xx_initial)/ncells        # Step size of spatial variable
nx = ncells+1                    # Number of points
domain = np.linspace(xx_initial+step_size/2.,xx_final,nx) # Computational Mesh of spatial variable.

# Pre allocation of Initial Conditions
m0 = zeros(nx)
n0 = zeros(nx)
u0 = zeros(nx)
halfcells = int(ncells/2)

IC = 2 # 6 Initial Conditoins cases are following
if IC == 1:
    print ("Initial Data Set 1")
    m0[:halfcells] = 1.5  ; m0[halfcells:] = 1.2;
    n0[:halfcells] = 1.8  ; n0[halfcells:] = 1.0;
    u0[:halfcells] = 2.0  ; u0[halfcells:] = 2.5;
    tEnd = 0.15;
elif IC == 2:
    print ("Initial Data Set 2")
    m0[:halfcells] = 0.8  ; m0[halfcells:] = 1.0;
    n0[:halfcells] = 1.0  ; n0[halfcells:] = 0.8;
    u0[:halfcells] = 1.2  ; u0[halfcells:] = 1.0;
    tEnd = 0.25;
elif IC == 3:
    print ("Initial Data Set 3")
    m0[:halfcells] = 2.0  ; m0[halfcells:] = 3.0;
    n0[:halfcells] = 2.0  ; n0[halfcells:] = 1.0;
    u0[:halfcells] = 2.5  ; u0[halfcells:] = 0.5;
    tEnd = 0.25;
elif IC == 4:
    print ("Initial Data Set 4")
    m0[:halfcells] = 3.2  ; m0[halfcells:] = 3.3;
    n0[:halfcells] = 2.5  ; n0[halfcells:] = 2.2;
    u0[:halfcells] = 2.9  ; u0[halfcells:] = 0.8;
    tEnd = 0.25;


q  = np.array([m0,n0,n0*u0])   # Vector of conserved variables

# Solver loop starts here
t  = 0 # Initial Time
lamda3 = max(u0+((k*gamma*(m0+n0)**gamma)/n0)**0.5) #Largest engenvalue of the system
dt=CFL*step_size/lamda3   # Using the system's largest eigenvalue


# Plot the Initial Data Set 
fig, axs = plt.subplots(3,sharex=True)
fig.suptitle('The Force Scheme - Initial Data Set '+str(IC)+'',fontsize=16)
axs[0].plot(domain, q[0],'k-')
axs[0].set_ylabel('m',fontsize=16)
axs[1].plot(domain, q[1],'r-')
axs[1].set_ylabel('n',fontsize=16)
axs[2].plot(domain, q[2],'b-')
axs[2].set_ylabel('u',fontsize=16)
axs[2].set_xlabel('x',fontsize=16)
fig.savefig("FORCE_Scheme_Initial_"+str(IC)+".jpeg", dpi=900)
plt.show()




while t < tEnd:
    
    # Making copy of conserved variable.
    Q = q.copy();
    
    # Primary variables
    m=Q[0];
    n=Q[1];
    u=Q[2]/n;
    
    
    # Flux vector of conserved properties
    F0 = np.array(m*u)
    F1 = np.array(n*u)
    F2 = np.array(n*u**2 +k*(m+n)**gamma)
    flux=np.array([ F0, F1, F2 ]);
    
    
    # Computing the Q(i-1) Q(i+1) F(Q(i+1)) F(Q(i-1))
    qm  = np.roll(Q, 1);
    qp  = np.roll(Q,-1);
    fm  = np.roll(flux, 1);
    fp  = np.roll(flux,-1);

    qpHalf = (qp+Q)/2. - dt/(2.*step_size)*(fp-flux)
    qmHalf = (qm+Q)/2. - dt/(2.*step_size)*(flux-fm)

    # Computing the Lax Friedrich Flux
    FqmLF = (fm + flux)/2.0  - step_size*(Q-qm)/(2.0*dt)
    FqpLF = (flux + fp)/2.0  - step_size*(qp-Q)/(2.0*dt)
    

    m=qpHalf[0];
    n=qpHalf[1];
    u=qpHalf[2]/n;
    
    F0 = np.array(m*u);
    F1 = np.array(n*u);
    F2 = np.array(n*u**2+k*(m+n)**gamma);
    FqpHalf=np.array([ F0, F1, F2 ]);
    
    m=qmHalf[0];
    n=qmHalf[1];
    u=qmHalf[2]/n;
    
    F0 = np.array(m*u);
    F1 = np.array(n*u);
    F2 = np.array(n*u**2+k*(m+n)**gamma);
    FqmHalf=np.array([ F0, F1, F2 ]);
    
    
    FM = (FqmHalf + FqmLF)/2.0
    FP = (FqpHalf + FqpLF)/2.0
    
    dF = FP - FM;
    
    q = Q-dt/step_size*dF;
    
    # Apply the Boundary Conditions (Periodic BC)
    q[:,0]=Q[:,0]; q[:,-1]=Q[:,-1];
   
  
    # Compute primary variables
    m=q[0];
    n=q[1];
    u=q[2]/n;

    # Update the time step
    lamda3 = max(u+((k*gamma*(m+n)**gamma)/n)**0.5);
    dt=CFL*step_size/lamda3;
    
    # Update time and iteration counter
    t=t+dt; 


fig, axs = plt.subplots(3,sharex=True)
fig.suptitle('The FORCE Scheme at t = '+str(round(t,2))+'',fontsize=16)
axs[0].plot(domain,m,'k-')
axs[0].set_ylabel('m',fontsize=16)
axs[1].plot(domain, n,'r-')
axs[1].set_ylabel('n',fontsize=16)
axs[2].plot(domain, u,'b-')
axs[2].set_ylabel('u',fontsize=16)
axs[2].set_xlabel('x',fontsize=16)
fig.savefig("FORCE_Scheme_Final_"+str(IC)+".jpeg", dpi=900)
plt.show()
\end{lstlisting}

\bibliographystyle{plain}

\begin{thebibliography}{99}

\bibitem{toro2013riemann}
E.~F. Toro.
\newblock \emph{Riemann Solvers and Numerical Methods for Fluid Dynamics: A Practical Introduction}.
\newblock Springer Science \& Business Media, 2013.

\bibitem{ChunShen}
C.~Shen.
\newblock The singular limits of solutions to the {R}iemann problem for the liquid--gas two-phase isentropic flow model.
\newblock \emph{Journal of Mathematical Physics}, 61(8):081502, 2020.

\bibitem{zeidan2007numerical}
D.~Zeidan, E.~Romenski, A.~Slaouti, and E.~F. Toro.
\newblock Numerical study of wave propagation in compressible two-phase flow.
\newblock \emph{International Journal for Numerical Methods in Fluids}, 54(4):393--417, 2007.

\bibitem{flaatten2006approximate}
T.~Fl{\aa}tten and S.~T. Munkejord.
\newblock The approximate {R}iemann solver of {R}oe applied to a drift-flux two-phase flow model.
\newblock \emph{ESAIM: Mathematical Modelling and Numerical Analysis}, 40(4):735--764, 2006.

\bibitem{zeidan2011riemann}
D.~Zeidan.
\newblock The {R}iemann problem for a hyperbolic model of two-phase flow in conservative form.
\newblock \emph{International Journal of Computational Fluid Dynamics}, 25(6):299--318, 2011.

\bibitem{zeidan2020high}
D.~Zeidan, L.~T. Zhang, and E.~Goncalves.
\newblock High-resolution simulations for aerogel using two-phase flow equations and {G}odunov methods.
\newblock \emph{International Journal of Applied Mechanics}, 12(05):2050049, 2020.

\bibitem{huang2018nonlinear}
F.~Huang, D.~Wang, and D.~Yuan.
\newblock Nonlinear stability and existence of vortex sheets for inviscid liquid-gas two-phase flow.
\newblock arXiv preprint arXiv:1808.05905, 2018.

\bibitem{ruan2016rectilinear}
L.~Ruan, D.~Wang, S.~Weng, and C.~Zhu.
\newblock Rectilinear vortex sheets of inviscid liquid-gas two-phase flow: linear stability.
\newblock \emph{Communications in Mathematical Sciences}, 14(3):735--776, 2016.

\bibitem{touma2014computations}
R.~Touma and D.~Zeidan.
\newblock On the computations of gas--solid mixture two-phase flow.
\newblock 2014.

\bibitem{evje2008global}
S.~Evje and K.~H. Karlsen.
\newblock Global existence of weak solutions for a viscous two-phase model.
\newblock \emph{Journal of Differential Equations}, 245(9):2660--2703, 2008.

\bibitem{00a3c9ca6ca82c1c78a8848bd4ac4a53}
M.~Hantke and C.~Matern.
\newblock The {R}iemann problem for a weakly hyperbolic two-phase flow model of a dispersed phase in a carrier fluid.
\newblock \emph{Quarterly of Applied Mathematics}, 78(3):431--467, 2020.

\bibitem{Share_it_1981185920-73243}
C.~Matern.
\newblock The {R}iemann problem for weakly hyperbolic two-phase flow model of a dispersed phase in a carrier fluid.
\newblock PhD thesis, Otto-von-Guericke-Universit{\"a}t Magdeburg, Fakult{\"a}t f{\"u}r Mathematik, 2022.

\bibitem{leveque1992numerical}
R.~J. LeVeque.
\newblock \emph{Numerical Methods for Conservation Laws}.
\newblock Springer, 1992.

\bibitem{leveque2002finite}
R.~J. LeVeque.
\newblock \emph{Finite Volume Methods for Hyperbolic Problems}.
\newblock Cambridge University Press, 2002.

\bibitem{Munkejord}
S.~T. Munkejord, S.~Evje, and T.~Fl{\aa}tten.
\newblock The multi-stage centred-scheme approach applied to a drift-flux two-phase flow model.
\newblock \emph{International Journal for Numerical Methods in Fluids}, 52:679--705, 2006.

\end{thebibliography}

\end{document}